\documentclass[aps,nofootinbib,showpacs,twocolumn,floatfix,superscriptaddress]{revtex4}
\usepackage{graphicx}
\usepackage{epsf,amsfonts,amssymb,amsbsy}

\begin{document}
\title{Scaling attractors for quintessence in flat universe with cosmological term}
\author{V.V.Kiselev}
 \affiliation{Russian State Research
Center ``Institute for High
Energy Physics'', 
Pobeda 1, Protvino, Moscow Region, 142281, Russia\\ Fax:
+7-4967-744937}
 \affiliation{Moscow Institute of Physics and
Technology, Institutskii per. 9, Dolgoprudnyi, Moscow Region,
141700, Russia}
\begin{abstract}
For evolution of flat universe, we classify late time and future
attractors with scaling behavior of scalar field quintessence in
the case of potential, which, at definite values of its parameters
and initial data, corresponds to exact scaling in the presence of
cosmological constant.
\end{abstract}
\maketitle

\section{Introduction}

Recent astronomical measurements of Super Novae Ia light curves
versus their red shifts (SNIa) \cite{snIa,deceldata,SNLS}, Cosmic
Microwave Background Radiation anisotropy (CMBR) by Wilkinson
Microwave Anisotropy Project (WMAP) \cite{wmap}, inhomogeneous
correlations of baryonic matter by Sloan Digital Sky Survey (SDSS)
and 2dF Galaxy Redshift Survey \cite{baryon} with a high precision
enforce the following picture of cosmology:
\begin{itemize}
    \item the Universe is flat,
    \item its evolution is consistently driven by cosmological
    constant $\Lambda$ and cold dark matter (CDM), that constitutes
    the $\Lambda$CDM model.
\end{itemize}
Irrespective of dynamical nature for such substances, that could
be different, at present any model of cosmology has to demonstrate
its tiny deviation from the $\Lambda$CDM evolution at late times,
i.e. the behavior of Hubble constant should scale extremely close
to
\begin{equation}\label{H2}
    H^2=H_0^2\left(\Omega_\Lambda+\frac{\Omega_M}{a^3}\right),
\end{equation}
with $H_0=H(t_0)$ denoting the present day Hubble constant,
$a=a(t)$ is the scale factor in the Friedmann--Robertson--Walker
metric
\begin{equation}\label{ds}
    {\rm d}s^2={\rm d}t^2-a^2(t)\,[{\rm d}r^2+r^2{\rm
    d}\theta^2+r^2\sin^2\theta{\rm d}\varphi^2],
\end{equation}
conveniently normalized by $a(t_0)=1$, so that $H=\dot a/a$ with
dot meaning the differentiation with respect to time $t$. The
fractions $\Omega_\Lambda$ and $\Omega_M$ represent the
cosmological term and pressureless matter including both baryons
and cold dark matter. In (\ref{H2}) we neglect contributions by
radiation fractions given by photons and neutrinos.

Dynamical models closely fitting the above behavior of Hubble
constant include a quintessence \cite{quint}, a scalar filed
$\phi$ with slowly changing potential energy $V(\phi)$ imitating
the contribution of cosmological constant (see recent review on
the reconstruction of dark energy dynamics in \cite{SS-rev}). In
present paper we find a potential of scalar field $\phi$ which
exactly reproduces the scaling behavior of Hubble constant in flat
universe with cosmological term\footnote{The other approach of
reconstructing an effective potential of variable cosmological
term by the given luminosity distance or inhomogeneity growth
factor was considered by A.Starobinsky in \cite{S-Ddelta}.}
(Section II). The potential is the square of hypersine with some
tuned values of normalization and slope. In Section III we study
the stability of scaling behavior versus the parameters of
potential in terms of autonomous system of differential equations
possessing critical points. We find the late time attractors that
further evolve to future attractors generically different from
those of late time. We discuss a physical meaning of attractors in
Section IV. In Conclusion we summarize our results.

\section{Exact solution with scaling behavior}

The evolution is described by following equations:
\begin{equation}\label{1}
    \left\{
    \begin{array}{l}\displaystyle
    H^2 = \frac{8\pi G}{3}\,\left(\rho_B+\frac{1}{2}\,{\dot
    \phi}^2+V(\phi)\right),\\[5mm]
    \displaystyle
    \ddot\phi+3 H \dot\phi+\frac{\partial
    V(\phi)}{\partial\phi} =  0,
    \end{array}
    \right.
\end{equation}
where $\rho_B$ is the energy density of baryotropic matter with
pressure $p_B=w_B\rho_B$, which satisfies the energy-momentum
conservation
\begin{equation}\label{matter}
    {\dot\rho}_B+3 H (\rho_B+p_B)=0,
\end{equation}
yielding the scaling behavior
\begin{equation}\label{matter2}
    \rho_B=\frac{\rho_0}{a^{3(1+w_B)}}.
\end{equation}
We suppose the following scale dependence of Hubble constant
\begin{equation}\label{h2}
    H^2=H_0^2\left(\Omega_\Lambda+\frac{\Omega_S}{a^{3(1+w_B)}}\right),
\end{equation}
where $\Omega_S$ denotes the present day fraction of substance
composed of baryotropic matter, cold dark matter and quintessence,
that could simulate dark substances ordinary introduced in the
standard consideration: the dark energy and dark matter. At late
times (or at present) $w_B=0$ corresponds to nonrelativisting
matter with negligibly small pressure (the dust), while $w_B=1/3$
stands for the radiation era of hot matter.

The critical density $\rho_c$ is defined by
\begin{equation}\label{h3}
    H_0^2=\frac{8\pi G}{3}\,\rho_c,
\end{equation}
so that the baryonic and dark matter fractions are presently given
by
\begin{equation}\label{matter3}
    \Omega_b=\frac{\rho_b}{\rho_c}\Big|_{t=t_0},\qquad
    \Omega_{DM}=\frac{\rho_{DM}}{\rho_c}\Big|_{t=t_0},
\end{equation}
while for brevity we put
\begin{equation}\label{Bb}
    \Omega_B=\Omega_b+\Omega_{DM},\quad\mbox{or}\quad
    \rho_B=\rho_b+\rho_{DM}.
\end{equation}

The scalar field density and pressure
\begin{equation}\label{scal2}
    \rho_\phi=\frac{1}{2}\,{\dot\phi}^2+V,\qquad
    p_\phi=\frac{1}{2}\,{\dot\phi}^2-V.
\end{equation}
determine the fraction
\begin{equation}\label{scal1}
    \Omega_\phi=\frac{1}{\rho_c}\,\rho_\phi\Big|_{t=t_0},
\end{equation}
and state parameter function
\begin{equation}\label{tw}
    w_\phi=\frac{p_\phi}{\rho_\phi}.
\end{equation}
The substance fraction is the sum of matter and field fractions
\begin{equation}\label{sub1}
    \Omega_S=\Omega_B+\Omega_\phi.
\end{equation}
We define the vacuum energy by
\begin{equation}\label{V_0}
    V_0=\rho_c\,\Omega_\Lambda,
\end{equation}
so that in flat universe
\begin{equation}\label{budget}
    \Omega_\Lambda+\Omega_S=1.
\end{equation}

Since we investigate the scaling behavior of functions homogeneous
with respect to scale factor $a$, it is convenient to introduce
the following variable
\begin{equation}\label{N}
    N=\ln a(t),
\end{equation}
so that the differentiation with respect to time denoted by dot is
reduced to the differentiation with respect to $N$ denoted by
prime,
\begin{equation}\label{n2}
    \dot \phi=\phi'\,H,\qquad \frac{\partial V}{\partial
    \phi}=\frac{V'}{\phi'}.
\end{equation}
Then, the equation of motion for $\phi$ is deduced to
\begin{equation}\label{phi}
    \frac{1}{2}\,\left(\{\phi'H\}^2\right)'+3\{\phi'H\}^2+V'=0.
\end{equation}

The consideration suggests that the potential scales
as\footnote{Symmetries of evolution equations were systematically
investigated in \cite{Marek}, wherein the authors found the
analogous behavior of energy density versus the scale factor and
used it for deriving the equation of state on the basis of
supernovae Ia data, while our goal is the potential itself.}
\begin{equation}\label{pot1}
    V=V_0+\frac{\widetilde \Omega_\phi}{a^{3(w_B+1)}}\,\rho_c,
\end{equation}
where $\widetilde \Omega_\phi$ is a constant. Therefore, the
derivative of potential scales, too,
\begin{equation}\label{pot2}
    V'=-3(w_B+1)\,(V-V_0).
\end{equation}
According to (\ref{phi}), we suggest
\begin{equation}\label{quad}
    3\{\phi'H\}^2=-c\,V',
\end{equation}
where the constant $c$ can be found from (\ref{phi}), since
\begin{equation}\label{quad2}
    (\{\phi'H\}^2)'=-\frac{c}{3}\,V'' =c(w_B+1)\,V',
\end{equation}
hence, (\ref{phi}) yields
\begin{equation}\label{phi2}
    \left\{\frac{1}{2}\,c(w_B+1)-c+1\right\}V'=0,
\end{equation}
that is satisfied at
\begin{equation}\label{c}
    c=\frac{2}{1-w_B}.
\end{equation}
One can easily get
\begin{equation}\label{dot2}
    \frac{1}{2}\,(\dot\phi)^2=\frac{1}{2}\,(\phi'H)^2=
    \frac{1+w_B}{1-w_B}\,\frac{\widetilde\Omega_\phi}{a^{3(w_B+1)}},
\end{equation}
so that
\begin{equation}\label{rho}
    \rho_\phi=\rho_c\,\left\{
    \Omega_\lambda+\frac{2}{a^{3(w_B+1)}}\,\frac{\widetilde\Omega_\phi}{1-w_B}
    \right\},
\end{equation}
that gives the relation
\begin{equation}\label{omega2}
    \Omega_\phi=\frac{2}{1-w_B}\,\widetilde\Omega_\phi.
\end{equation}

In order to restore the potential yielding the scaling behavior,
we have to resolve (\ref{quad}) making use of (\ref{pot1}),
(\ref{c}), i.e.
\begin{equation}\label{Vphi}
    (\phi')^2=\frac{2\rho_c\widetilde\Omega_\phi}{H^2}\,\frac{1+w_B}{1-w_B}\,
    \frac{1}{a^{3(w_B+1)}},
\end{equation}
where
$$
    H^2=\frac{8\pi
    G}{3}\,\rho_c\left\{\Omega_\Lambda+\frac{1}{a^{3(w_B+1)}}\left(
    \Omega_B+\frac{2\widetilde\Omega_\phi}{1-w_B}\right)\right\}.
$$
The integration straightforwardly yields
\begin{equation}\label{phi3}
    \frac{\lambda}{2}\,\kappa\,(\phi-\phi_\star)
    =\mbox{arcsinh}\sqrt{\frac{1}{a^{3(w_B+1)}}\,\frac{\Omega_S}{\Omega_\Lambda}},
\end{equation}
where $\kappa^2=8\pi G$, and
\begin{equation}\label{lam1}
    \lambda=\sqrt{\,3(1+w_B)\,\frac{\Omega_S}{\Omega_\phi}}.
\end{equation}
For brevity of formulae we put the integration constant
$\phi_\star=0$ with no lose of generality. Then,
\begin{equation}\label{pot-end}
    V=V_0\left(
    1+\frac{\widetilde\Omega_\phi}{\Omega_S}\,
    \sinh^2\left\{\frac{\lambda}{2}\,\kappa\,\phi\right\}
    \right),
\end{equation}
where
\begin{equation}\label{frac}
    \frac{\widetilde\Omega_\phi}{\Omega_S}=\frac{3}{2}\,\frac{1-w_B^2}{\lambda^2},
\end{equation}
while
\begin{equation}\label{HH2}
    H^2=H_0^2\Omega_\Lambda
    \cosh^2\left\{\frac{\lambda}{2}\,\kappa\,\phi\right\}.
\end{equation}
So, at the present day we have
$\cosh^2\left\{\frac{\lambda}{2}\,\kappa\,\phi_0\right\}=1/\Omega_\Lambda$.

Summarizing the result, we emphasize that there is the exact
solution for the scalar field potential (\ref{pot-end}), which
reproduce the scaling behavior of Hubble constant in the evolution
of flat universe in presence of cosmological constant.

The form of potential differs from the case of zero cosmological
constant, where the potential is the exponent as was studied for
the scalar field with the standard kinetic term in
\cite{Wetterich,CLW,FJ,Alb_Skordis}, while the consideration for
the general scalar field was developed in \cite{Tsuji,GWZ}. One
can easily notice that the present derivation is consistent with
results concerning for the case of zero cosmological constant.
Indeed, the integration of (\ref{Vphi}) with the Hubble rate at
$\Omega_\Lambda=0$ straightforwardly gives the field proportional
to the logarithm of scale factor, $\phi\varpropto \ln a$, that
makes the scaling behavior of potential $V\varpropto
\exp\{\tilde\lambda\kappa\phi\}$.

The potential derived can be represented in the form
\begin{equation}\label{pot-end2}
    V=V_0+\frac{1}{2}\,\widetilde V_0\big(
    \cosh\{\lambda\kappa\phi\}-1\big),
\end{equation}
at $\widetilde V_0=3 V_0(1-w_B^2)/2\lambda^2$. So, function
(\ref{pot-end2}) is composed by the sum of two exponential
potentials with the opposite slopes and the constant positive
shift of minimum. Such kind of potentials was investigated
recently.

In review \cite{SS-00} authors presented the exact solution for
constant parameter of state for the dark energy in the presence of
dust-like dark matter, but not adding the cosmological constant,
which is imitated by the dark energy instead. In \cite{SahniWang}
the evolution of scalar field with the hyper-cosine potential was
studied in the case of zero cosmological constant: the exponential
form found to be dominant at early times, while the square term
did significant at late times. It is clear that the late time
dynamics essentially changed by the presence of cosmological term,
of course.

Various aspects of cosmological picture due to a potential given
by a power of hyper-sine was investigated in \cite{UrenaMatos}.
The tracker properties of such the potentials were stressed.

Another approach was presented in \cite{GF,SenSethi,RSPC,RSPCC},
where authors fixed the scaling behavior of Hubble constant to
find exact solutions for the scale factor $a(t)$ in order to study
characteristics of universe evolution. The consideration of
\cite{GF} recovers the scale factor behavior in the case of
$\Lambda$CDM, however, the authors did not address the question on
the scalar field potential reproducing such the scaling. This
question was investigated in \cite{SenSethi}, where the potential
with the same form of (\ref{pot-end2}) was deduced in a particular
case of $\Omega_S/\Omega_\Lambda=1/\sinh^2 1$ and $w_B=0$. At this
choice $\Omega_\Lambda=\sinh^2 1/(1+\sinh^2 1)\approx 0.58$, which
is in contradiction with the recent measurements \cite{wmap}
yielding $\Omega_\Lambda=0.766\pm 0.035$. A cosmological
exploration of potential composed by sum of two exponents with
opposite slopes but generically different normalization factors
and a negative shift of minimum was investigated in
\cite{RSPC,RSPCC} by the same method of exact time dependence. The
authors found an oscillation of $a(t)$ around some scaling
dependence with $w_\phi$ oscillating within $[-1;+1]$.

In \cite{BarreiroCN} the sum of two exponents with identical
normalization factors but slopes, which can be different, was
considered at first. The late time behavior of scalar field energy
scales both the radiation and dust, while near the present and
future the state parameter $w_\phi$ has decaying vibrations around
$-1$. Such picture differs from that of \cite{RSPC,RSPCC}.
Questions are the followings: \textit{i)} What is a reason for the
difference? \textit{ii)} What can we say about a stability of late
time and future scaling? \textit{iii)} Does presented exact
scaling solution corresponds to fine tuned values of normalization
and slope? These questions were not investigated in references
mentioned. We address them in Section III.

\section{Attractors}

Let us consider the evolution of flat universe in presence of
scalar field with potential
\begin{equation}\label{a1}
    V=V_0+\widetilde V_0\sinh^2\left\{\frac{\lambda}{2}\,
    \kappa\phi\right\},
\end{equation}
where $V_0$, $\widetilde V_0$ and $\lambda$ are free parameters,
which are not fixed by values in (\ref{lam1}), (\ref{pot-end}).
For definiteness we put all parameters to be positive: $V_0>0$,
$\widetilde V_0>0$, $\lambda>0$, while the consideration for cases
of negative values can be rather straightforwardly obtained from
the formulae below. The Hubble constant is given by
$$
    H^2=\frac{\kappa^2}{3}\left(
    \rho_B+\frac{1}{2}\,(\dot\phi)^2+V_0+\widetilde V_0\sinh^2
    \left\{\frac{\lambda}{2}\,
    \kappa\phi\right\}\right),
$$
so that we introduce quantaties $U_0$ and $U$, so that
\begin{equation}\label{U0}
    U_0^2=\frac{\kappa^2}{3}\,V_0,\qquad H^2 = U^2+U_0^2.
\end{equation}
Then, the phase space of system is described by dimensionless
variables
\begin{equation}\label{vary}
    x=
    \frac{\kappa}{\sqrt{6}}\,\frac{\phi'H}{U},\quad
    y=
    \frac{\kappa}{\sqrt{3}}\,\frac{\sqrt{V-V_0}}{U},\quad
    v=
    \frac{U_0}{U},
\end{equation}
while for convenience we introduce
\begin{equation}\label{z}
    z=-\frac{1}{\sqrt{V-V_0}}\,\frac{\partial
    V}{\partial\phi}\,\frac{\sqrt{2}}{H}.
\end{equation}
This choice of variables follows the observation of scaling in
previous section: the kinetic energy and potential each scale like
the Hubble constant squared after the subtraction of term caused
by the cosmlogical constant, while the derivative of potential
with respect to the field scales like the Hubble constant squared
itself.

The definition of $U^2$ implies
\begin{equation}\label{xy}
    x^2+y^2=1-\frac{\kappa^2}{3}\,\frac{\rho_B}{U^2},
\end{equation}
which yields the constraint
\begin{equation}\label{xy2}
    x^2+y^2\leqslant 1.
\end{equation}
The dynamical state parameter of field is determined by
\begin{equation}\label{ww}
    {\widetilde w}_\phi=\frac{p_\phi+V_0}{\rho_\phi-V_0}=\frac{x^2-y^2}{x^2+y^2}.
\end{equation}

In addition, the equations of motion produce relations
\begin{equation}\label{HHprime}
    \dot H=H'H=-(1-w_B)\rho_B-(\dot\phi)^2,\qquad U'U=H'H.
\end{equation}

The differentiation gives the autonomous system of equations
\begin{equation}\label{sys}
    \begin{array}{rcl}
      x' & = & -3x+\frac{1}{2}\,yz+\frac{3}{2}\,x\,c(x,y),\\[4mm]
      y' & = & -\frac{1}{2}\,xz+\frac{3}{2}\,y\,c(x,y),\\[4mm]
      (1+v^2)z' & = & 3\lambda^2xy-\frac{3}{2}\,z\,c(x,y),\\[4mm]
      v' & = & \frac{3}{2}\,v\,c(x,y),
    \end{array}
\end{equation}
where $c(x,y)=(1+w_B)(1-x^2-y^2)+2x^2$.  The quantity $z$ is
strictly constrained by the condition
\begin{equation}\label{z-cond}
    \frac{z^2}{6\lambda^2}\,(1+v^2)-y^2=\frac{\widetilde
    V_0}{V_0}\,v^2,
\end{equation}
which is the direct consequence of hyper-trigonometry:
$\cosh^2q-\sinh^2q=1$. This constraint makes system (\ref{sys})
overdefined, since $z$ is completely given by $y$, $v$ and
parameter $\xi^2=\widetilde V_0/V_0$. Nevertheless, the system
allows us to get a complete analysis of critical points in the
simplest manner. What is of our interest? It is the projection of
trajectory in the $\{x,y,v\}$ 3D-space to the 2D-plane of
$\{x,y\}$.

\subsection{Late times} At present, the cosmological constant
makes a significant contribution to the Hubble constant, i.e.
$v=U_0/U \sim 1$. At late times of evolution just before the
present, we put $v\ll 1$. Then, $z$ can be excluded by
\begin{equation}\label{z-cond2}
    z=\lambda\sqrt{6}\,y.
\end{equation}
This limit means that the cosmological constant can be neglected,
while the field has a large value, so that the hyper-sine can be
approximated by a single exponent. Therefore, at late times we
arrive to the analysis of exponential potentials given in
\cite{CLW}. Indeed, under (\ref{z-cond2}) system (\ref{sys}) is
reduced to the system for the exponential potential. The analysis
of critical point in \cite{CLW} gave the following physically
meaningful properties: irrelevant of normalization of potential
$\widetilde V_0$ there are stable scaling attractors in the plane
of $\{x,y\}$; these attractors appear at $\lambda^2> 3(1+w_B)$, so
that $\Omega_S/\Omega_\phi=\lambda^2/3(1+w_B)>1$ and $w_\phi=w_B$.
The attractor is the stable node at
$\lambda^2<24(1+w_B)^2/(7+9w_B)$, otherwise it is the stable
spiral focus. The position of attractors are given by
\begin{equation}\label{critic1}
    x_c=\sqrt{\frac{3}{2}}\,\frac{1+w_B}{\lambda},\qquad
    y_c=\sqrt{\frac{3(1-w_B^2)}{2}}\,\frac{1}{\lambda},
\end{equation}
that fixes z according to (\ref{z-cond2}), i.e.
$z_c=3\sqrt{1-w_B^2}$.

Thus, at late times just before the present, the quintessence
follows the scaling behavior independently of its initial
conditions.

\subsection{Future} Since function $c(x,y)$ takes positive values
at $x\neq 0$, $y\neq 0$, i.e. at presence of scalar field,
quantity $v$ grows in accordance with its differential equation in
(\ref{sys}), of course. Hence, in future we get $v\gg 1$. Then,
(\ref{z-cond}) yields
\begin{equation}\label{z-cond3}
    z\equiv z_\star=\lambda\sqrt{\frac{6\widetilde V_0}{V_0}}=\lambda\xi\sqrt{6},
\end{equation}
i.e., $z$ is frozen at $z=z_\star$. Therefore, we get the system
for the plane $\{x,y\}$ in (\ref{sys}) with $z_\star$ being the
external parameter. The critical points are posed at the following
sets:

I. Scalar field is absent,
\begin{equation}\label{critic2-0}
    \mbox{(i):}\quad x_\star=0,\qquad
    y_\star=0,
\end{equation}
so that the linearized equations in vicinity of critical point,
i.e. with $x=x_\star+\bar x$, $y=y_\star+\bar y$, result in the
system
\begin{equation}\label{sys-i}
    \left(%
    \begin{array}{c}
  \bar x' \\
  \bar y' \\
    \end{array}%
    \right)=\hat B\cdot
        \left(%
    \begin{array}{c}
  \bar x \\
  \bar y \\
    \end{array}%
    \right)
\end{equation}
with the matrix
\begin{equation}\label{B-i}
    \hat B=\frac{1}{2}\left(%
    \begin{array}{cc}
  3(w_B-1) & z_\star \\[2mm]
  -z_\star & 3(w_B+1) \\
    \end{array}%
    \right)
\end{equation}
having eigenvalues
\begin{equation}\label{nu-i}
    \begin{array}{l}
    \nu_1=\frac{1}{2}(3w_B-\sqrt{9-z_\star^2}),\\[3mm]
    \nu_2=\frac{1}{2}(3w_B+\sqrt{9-z_\star^2}),
    \end{array}
\end{equation}
that implies the critical point corresponds to instability due to
$\Re\mathfrak{e}\,\nu_2>0$ at $w_B>0$. Anyway, critical point
(\ref{critic2-0}) is a saddle at $|z_\star|<3$ and $w_B=0$, which
is of practical interest to look at the present time and future
universe. We notice that actually the baryonic matter has a small
pressure, which can be neglected in the universe evolution, i.e.
$w_B\to+0$. At $|z_\star|>3$ eigenvalues (\ref{nu-i}) satisfy
$\nu_1=\nu_2^*$, and we have unstable focus at $w_B>0$ or center
at $w_B\equiv 0$, while at $w_B<0$ the focus becomes stable.

II. Critical points of general position are given by
\begin{equation}\label{critic2-1}
    \mbox{(ii):} \left\{
    \begin{array}{l}
    x_\star=\frac{1}{\sqrt{6}}\,\sqrt{3-\sqrt{9-z_\star^2}},\\[4mm]
    y_\star=\frac{1}{\sqrt{6}}\,\sqrt{3+\sqrt{9-z_\star^2}},
    \end{array}
    \right.
\end{equation}
with additional symmetry over the following permutations:
$\mathcal{A}\mapsto$ $\{x_\star\leftrightarrow -x_\star$ and
$y_\star\leftrightarrow -y_\star\}$, $\mathcal{B}\mapsto$
$\{x_\star\leftrightarrow y_\star\}$, $\mathcal{C}\mapsto$ $\{$the
product of operations $\mathcal{A}$ and $\mathcal{B}\}$. So,
taking into account the symmetry, (\ref{critic2-1}) covers 4
related sets. The action of permutation $\mathcal{A}$ conserves
the stability properties, while action of $\mathcal{B}$ changes
them. This fact becomes clear, for instance, at $|z_\star|<3$.
Indeed, altering the sign for $x$ implies the interchange of two
equivalent branches in the potential according to
$\phi\leftrightarrow-\phi$ or formal altering the sign of quantity
$U$, both of which change the sign of $y$, too. So, action
$\mathcal{A}$ cannot influence the stability properties. The
action of $\mathcal{B}$ interchanges fractions of kinetic and
potential energies in the energy density, that can be physically
essential.

The analysis of linear stability involves matrices for
(\ref{critic2-1}) including action of symmetry $\mathcal{A}$ and
for (\ref{critic2-1}) converted by action of symmetry
$\mathcal{B}$, with corresponding eigenvalues $\{\nu\}$,
\begin{widetext}
\begin{equation}\label{B-iia}
    \hat B_a=\frac{1}{2}\left(%
    \begin{array}{cc}
  w_B(\sqrt{9-z_\star^2}-3)-2\sqrt{9-z_\star^2} & -z_\star w_B \\[2mm]
  -z_\star w_B & -w_B(\sqrt{9-z_\star^2}+3)-2\sqrt{9-z_\star^2} \\
    \end{array}%
    \right),\quad
    \quad
    \left\{
    \begin{array}{l}
    \nu^a_1=-3w_B-\sqrt{9-z_\star^2},\\[3mm]
    \nu^a_2=-\sqrt{9-z_\star^2},
    \end{array}
    \right.
\end{equation}
\begin{equation}\label{B-iib}
    \hat B_b=\frac{1}{2}\left(%
    \begin{array}{cc}
  -w_B(\sqrt{9+z_\star^2}+3)+2\sqrt{9-z_\star^2} & -z_\star w_B \\[2mm]
  -z_\star w_B & w_B(\sqrt{9-z_\star^2}-3)+2\sqrt{9-z_\star^2} \\
    \end{array}%
    \right),\quad
    \quad
    \left\{
    \begin{array}{l}
    \nu^b_1=-3w_B+\sqrt{9-z_\star^2},\\[3mm]
    \nu^b_2=+\sqrt{9-z_\star^2}.
    \end{array}
    \right.
\end{equation}
\end{widetext}
For real critical points, i.e. at $|z_\star|<3$, and at
$w_B\geqslant 0$, set (\ref{critic2-1}) is stable node (both
eigenvalues are negative), while set (\ref{critic2-1}) affected by
action $\mathcal{B}$ gives saddle or unstable node depending on
the sign of eigenvalue $\nu^b_1$, i.e. a balance between values of
$3w_B$ and $\sqrt{9-z_\star^2}$.

\begin{figure*}[ht]
  \includegraphics[width=7.25cm]{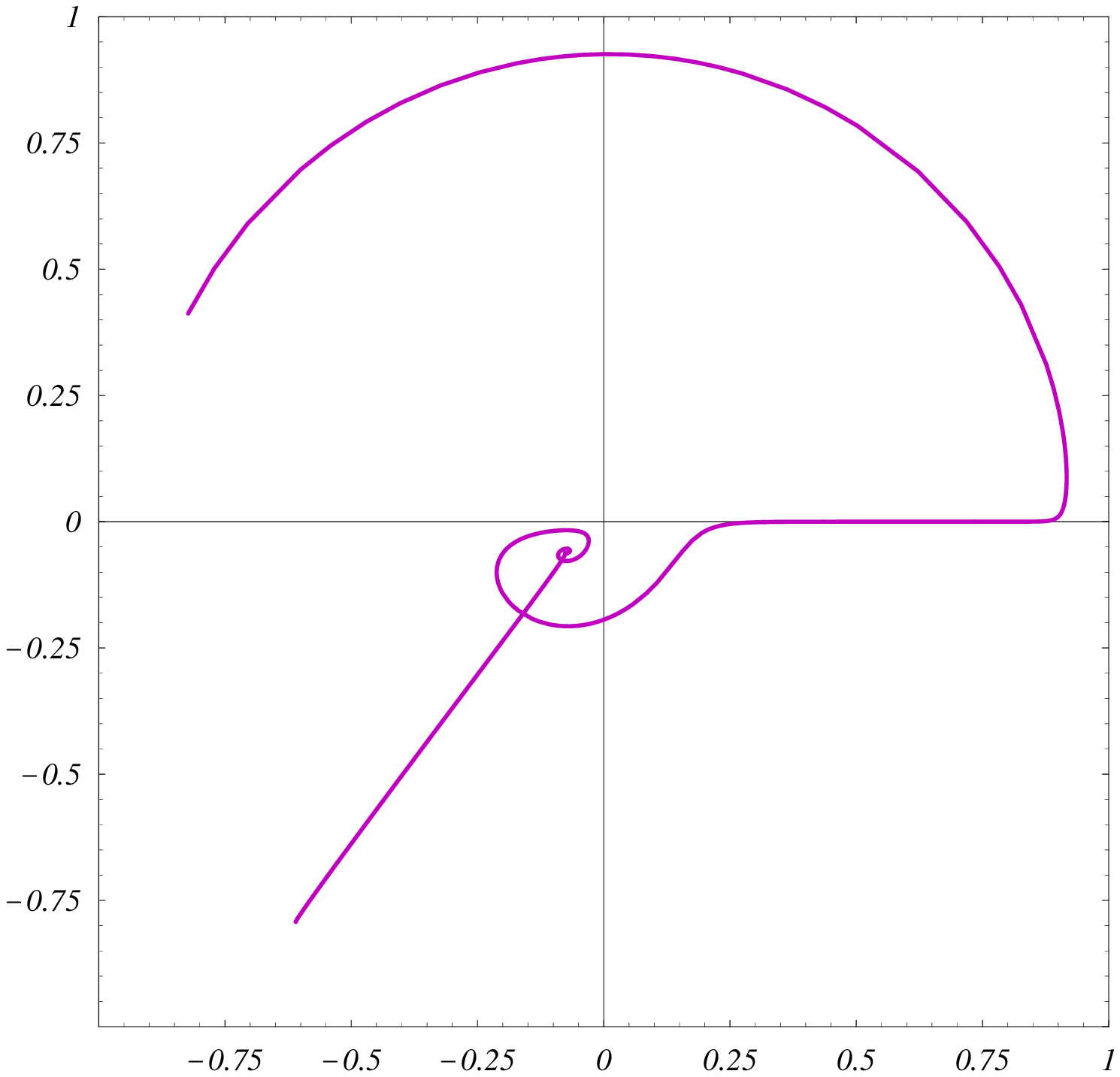}
  \hspace*{-1cm}\raisebox{1cm}{a)}
  \hspace*{7.5mm}
  \includegraphics[width=7.25cm]{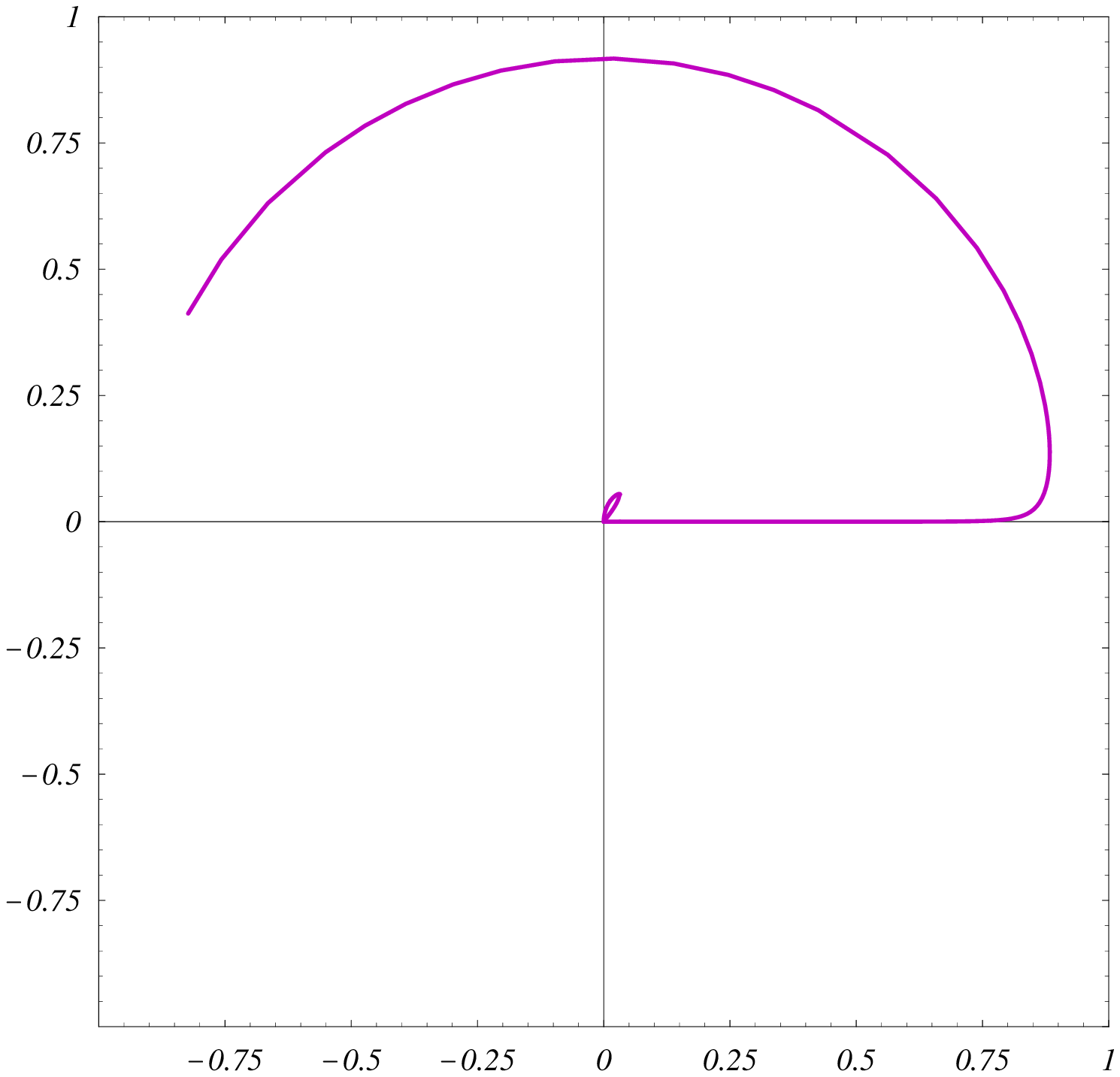}
  \hspace*{-1cm}\raisebox{1cm}{b)}\\[2mm]
  \includegraphics[width=7.25cm]{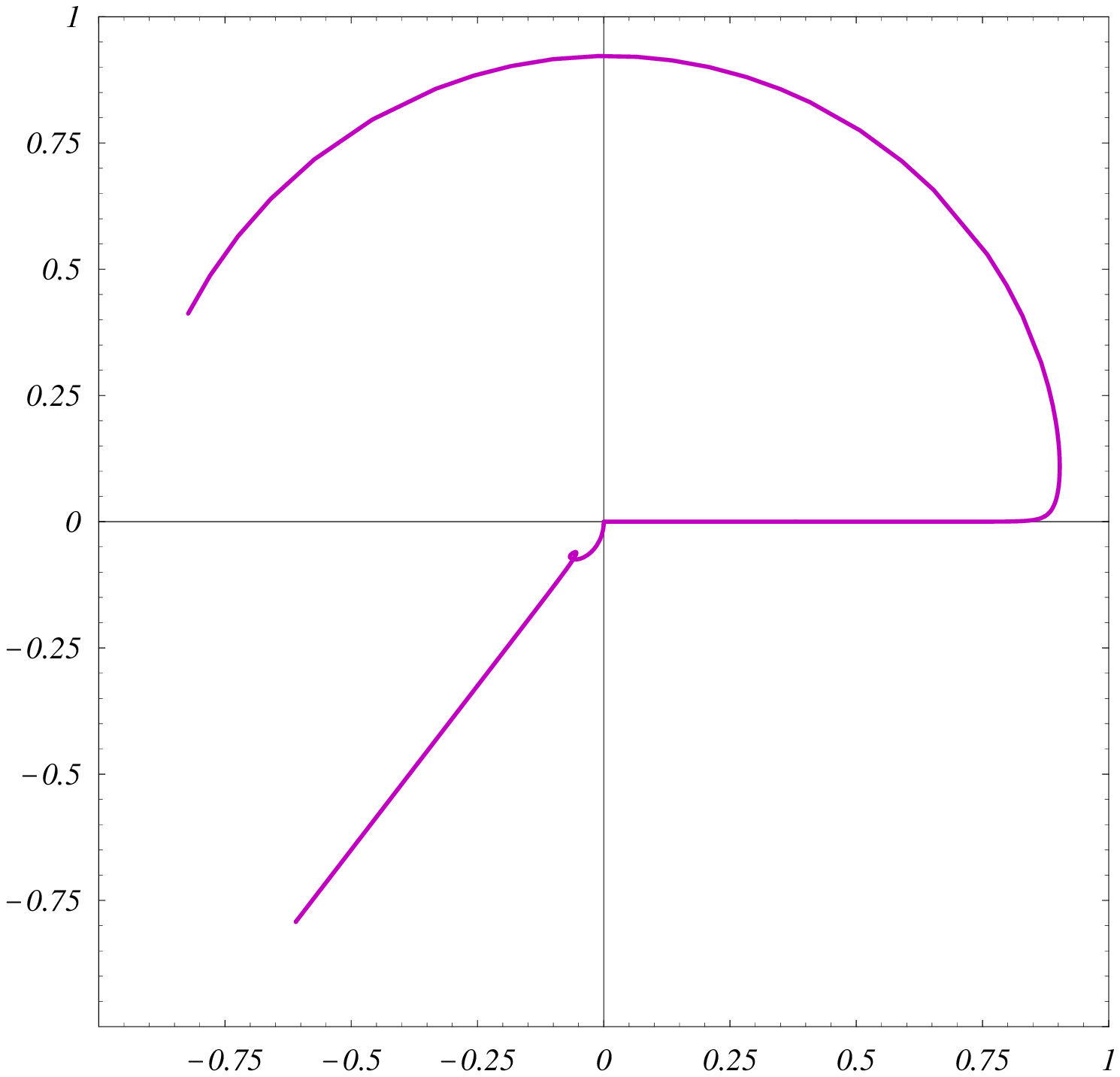}
  \hspace*{-1cm}\raisebox{1cm}{c)}
  \hspace*{7.5mm}
  \includegraphics[width=7.25cm]{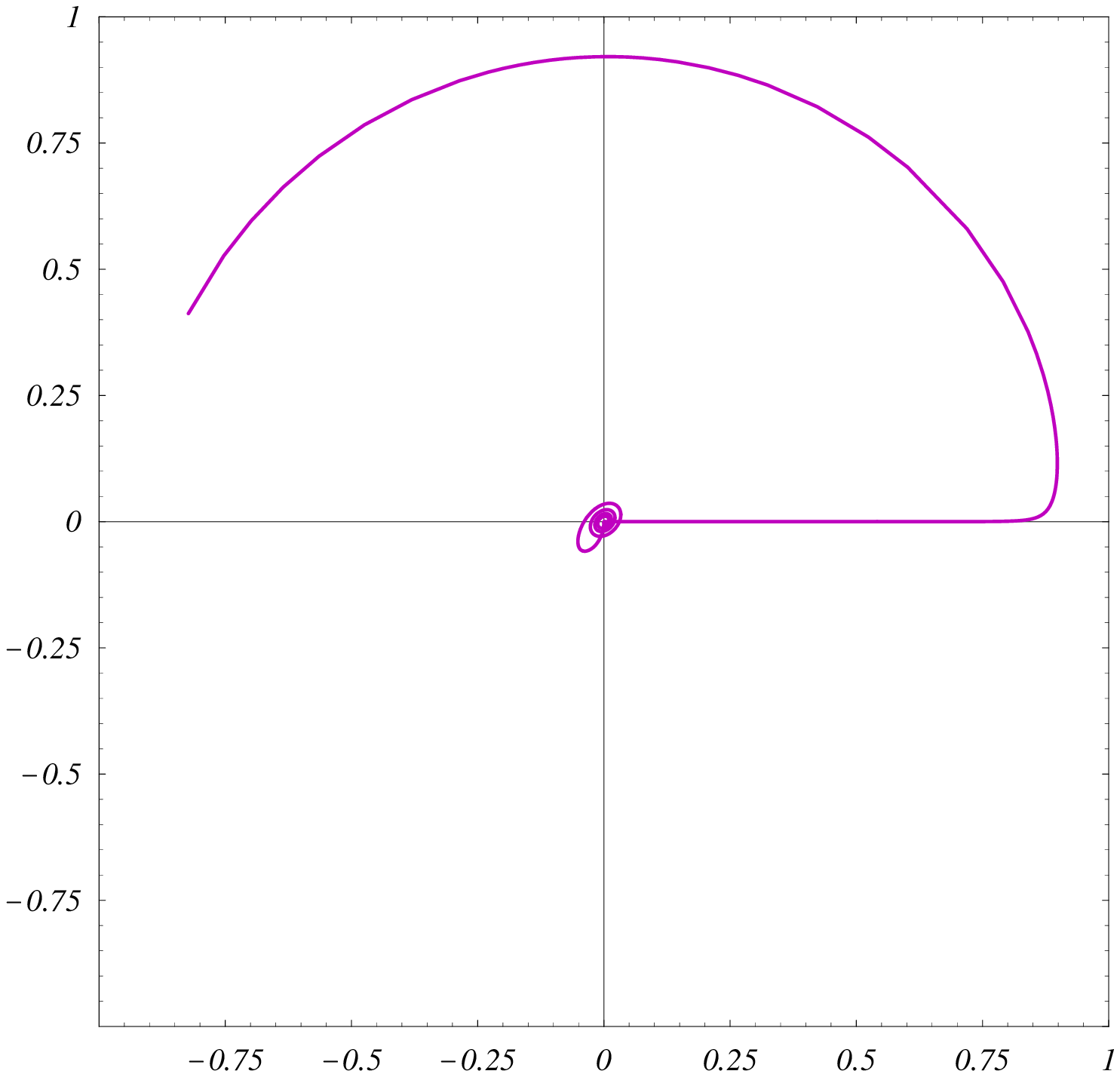}
  \hspace*{-1cm}\raisebox{1cm}{d)}\\[2mm]
  \includegraphics[width=7.25cm]{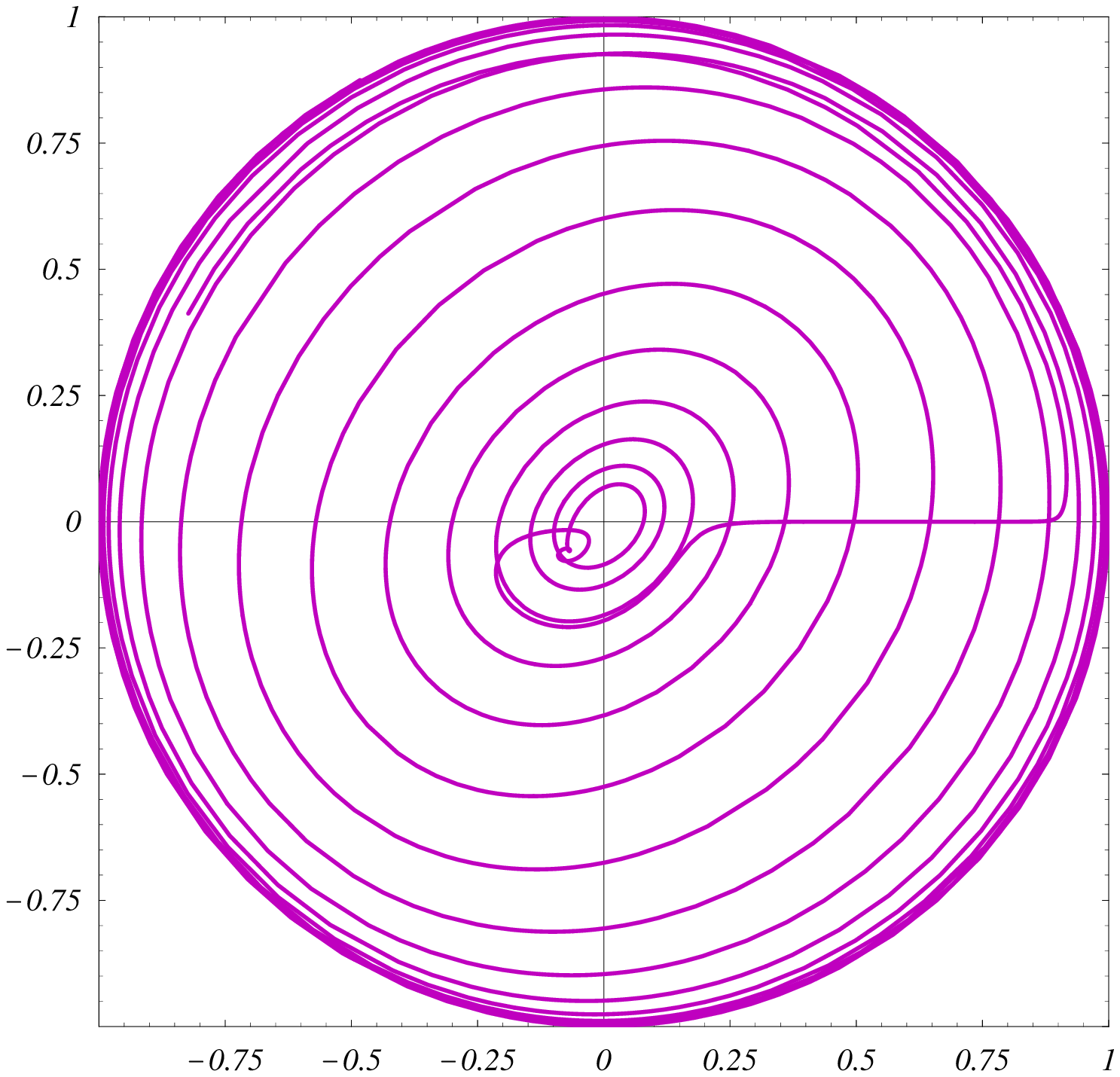}
  \hspace*{-1cm}\raisebox{1cm}{e)}
  \hspace*{7.5mm}
  \includegraphics[width=7.25cm]{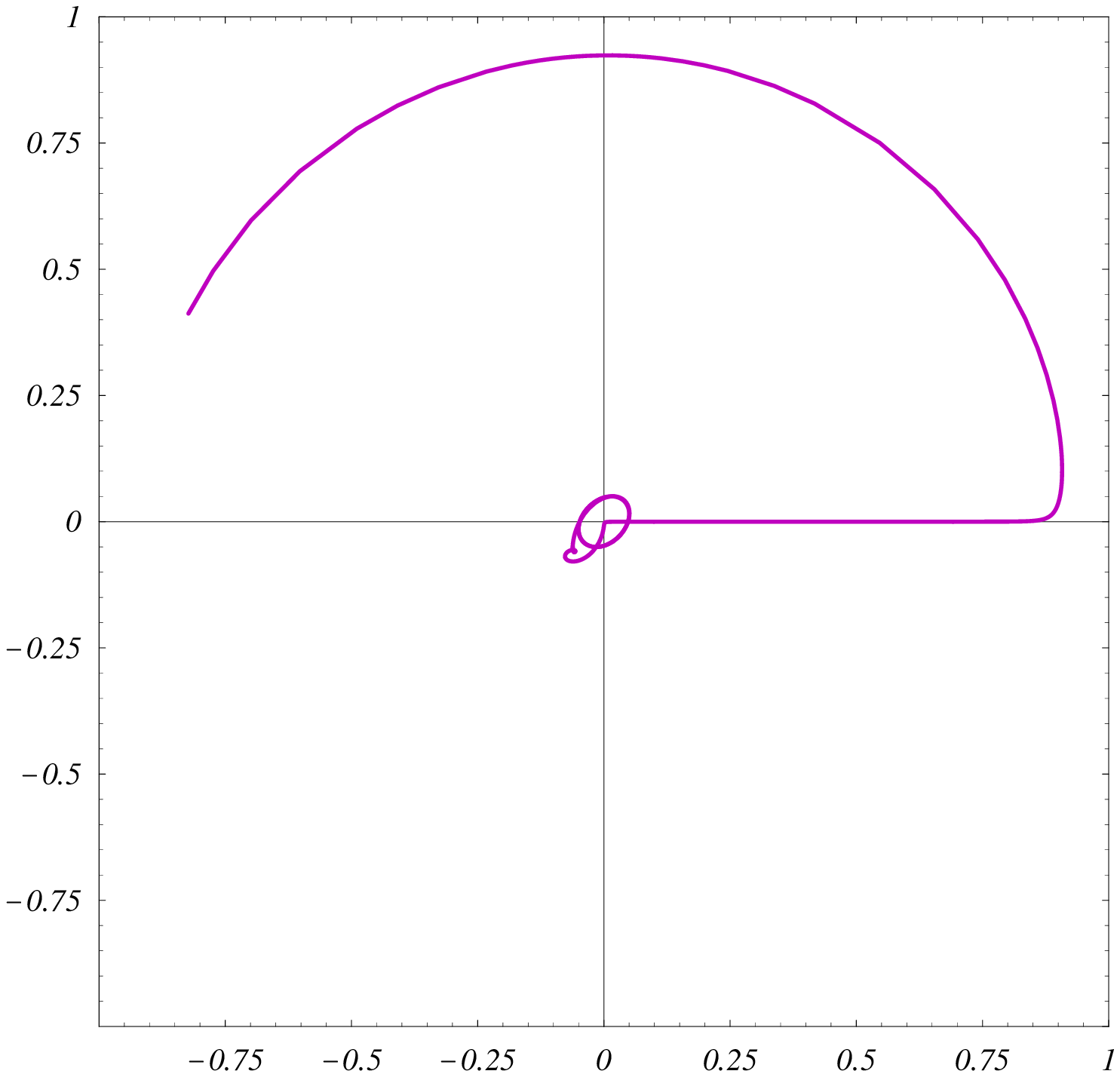}
  \hspace*{-1cm}\raisebox{1cm}{f)}
  \caption{The evolution of quintessence projected to the $\{x,y\}$ plane
  at slope $\lambda=20$ with various parameters of potential $z_\star$ and
  matter state equations $w_B$ as described in the text.}\label{phase-plot}
\end{figure*}

At $|z_\star|>3$ critical points (\ref{critic2-1}) take complex
values, while matrices satisfy $\hat B_a=\hat B_b^*$, i.e. they
are complex conjugate to each other as well as
$x_\star=y_\star^*$. The same is valid for eigenvectors:
\begin{equation}\label{eigenV}
    \boldsymbol e^a_1=
    \left(%
\begin{array}{c}
  3-\sqrt{9-z_\star^2}    \\
  z_\star \\
\end{array}%
\right),\quad
    \boldsymbol e^a_2=
    \left(%
\begin{array}{c}
  3+\sqrt{9-z_\star^2}    \\
  -z_\star \\
\end{array}%
\right),
\end{equation}
while $\boldsymbol e^b_{1,2}=\{\boldsymbol e^a_{1,2}\}^*$. Basis
(\ref{eigenV}) is orthonormal:
$$
    \boldsymbol e^a_i\cdot\boldsymbol e^a_j=\delta_{ij}.
$$
The linearized system has solutions
\begin{equation}\label{sol1}
    \left(%
\begin{array}{c}
  x \\
  y \\
\end{array}%
\right)=
    \left(%
\begin{array}{c}
  x_\star \\
  y_\star \\
\end{array}%
\right) +\sum\limits_{j=1}^2\boldsymbol e^a_j\,u_j\,{\rm
e}^{\nu^a_j N},
\end{equation}
where $u_j$ stand for some initial data. Solutions (\ref{sol1})
are complex-valued. By our investigation, (\ref{sol1}) are
irrelevant to physical quantities in question at $|z_\star|>3$.

III. Tuned scaling appears at the special value of parameter
$z_{\scriptscriptstyle T}=3\sqrt{1-w_B^2}$. Then, there is the
\textit{critical line}
\begin{equation}\label{line1}
    \mbox{(iii): ~ }y_\star=x_\star\sqrt{\frac{1-w_B}{1+w_B}},
\end{equation}
which is \textit{$\lambda$-invariant}, i.e. independent of slope
in the potential. It is spectacular that tuned point given by
$z_{\scriptscriptstyle T}=z_c$, $x_\star=x_c$, $y_\star=y_c$
corresponds to the exact scaling solution found in Section II.

At (\ref{line1}) the linear analysis of perturbations gives two
eigenvalues: $\nu_1=0$ and $\nu_2=2w_B$, so that zero eigenvalue
corresponds to the line itself, while the positive one indicates
instability at $w_B>0$ in vicinity of the line as a whole. This
fact is in agreement with the study of critical points in the case
II, since (\ref{line1}) contains the point of (\ref{critic2-1}) at
$w_B<0$ as well as it does the point at $w_B>0$ after the action
of symmetry $\mathcal{B}$, so we correspondingly get eigenvalues
(\ref{B-iia}) and (\ref{B-iib}) at $z_\star=z_{\scriptscriptstyle
T}$ yielding the same result above. Therefore, at $w_B>0$ the
scaling solution of section II is unstable in future, while it
does at $w_B<0$.

IV. The boundary circle
\begin{equation}\label{boundary1}
    \mbox{(iv): ~ }x^2+y^2=1
\end{equation}
is conserved by the autonomous system. This fact implies that at
$|z_\star|>3$ and $w_B>0$, when there are no stable critical
points, the system approaches the boundary circle in future.

Summarizing the results, we have found that in the presence of
cosmological constant the scalar field with the hyper-cosine
potential approaches the stable attractor at late times just
before the present day (the time of cosmological constant becomes
visible) at the potential slope $\lambda^2>3 (1+w_B)$, so that the
energy of field scales like that of matter with the state
parameter $w_B$, while the fraction of field energy depends on the
slope value. However, this attractor exhibits the strange behavior
in future, i.e. under the dominance of cosmological term. Then,
the balance of remnant energy depends on the parameter $z_\star$
determined by two quantities: i) the weight of potential
normalization with respect to the energy density due to the
cosmological constant and ii) the slope, in accordance with
(\ref{z-cond3}). So, at $|z_\star|<3$ and $w_B\geqslant0$ the
dynamical part of scalar field energy reaches the other scaling
attractor, which dominates over the matter in accordance with
(\ref{B-iia}) (see Fig. \ref{phase-plot}a) and gets the state
parameter
\begin{equation}\label{tww}
    \widetilde w_\phi=-\sqrt{1-\frac{z_\star^2}{9}}.
\end{equation}

At $|z_\star|<3$ and $w_B<0$ we get an interplay of two attractors
with (\ref{nu-i}) and (\ref{B-iia}): at
$3w_B+\sqrt{9-z_\star^2}<0$ the scalar field relaxes at the
minimum of kinetic and potential energy (\ref{critic2-0}) (see
Fig. \ref{phase-plot}b), while, otherwise, at
$3w_B+\sqrt{9-z_\star^2}>0$ the scalar field dominates over the
matter at the same point of (\ref{B-iia}) and (\ref{tww}) (see
Fig. \ref{phase-plot}c).

At $|z_\star|>3$ we get three cases: i) at $w_B>0$ the
quintessence in future vibrates at the boundary circle as the
limit cycle (see Fig. \ref{phase-plot}e), ii) at $w_B=0$ the field
cycled around the center being the point of minimal kinetic and
potential energy (see Fig. \ref{phase-plot}f), iii) at $w_B<0$ the
field relaxes to the minimum (see Fig. \ref{phase-plot}d).

The evolution kinds of scalar field in the plane of $\{x,y\}$ are
illustrated in Fig. \ref{phase-plot} at $\lambda=20$ and various
sets of $z_\star$ and $w_B$: at $z_\star=2.9$ we put $w_B=0.2$ in
Fig. \ref{phase-plot} a), $w_B=-0.5$ in b), and $w_B=-0.2$ in c),
while at $z_\star=10$ we put $w_B=-0.2$ in Fig. \ref{phase-plot}
d), $w_B=0.2$ in e), and $w_B=0$ in f). At all of tries the
evolution starts at $x_0=-0.8$, $y_0=0.4$, and trajectories move
clockwise. We certainly see that trajectories approaches the late
time attractor at appropriate $\{x_c,y_c\}$, which are negative in
all cases except b), when they are positive. The future attractors
in Fig. \ref{phase-plot} a) and c) are posed at
$\{x_\star,y_\star\}$ with opposite, negative, sign of
(\ref{critic2-1}). The attractors at b) and d) stand in the
minimum, while limit cycles of e) and f) are at the border and
around the minimum, correspondingly.

Thus, in the linear analysis we have classified the late time and
future attractors for the quintessence with the specified kind of
potential relevant to the case of nonzero cosmological constant.

However, this analysis falls in special degenerate case of
$z_\star=3$, $w_B=0$.

\subsection{Degenerate case}

At $z_\star=3$, $w_B=0$ and $v\gg 1$ the analysis of future
evolution becomes nonlinear, since, after the transformation to
variables $\sigma=x^2+y^2$ and $\tau=x/y$, autonomous system
(\ref{sys}) is reduced to
\begin{equation}\label{sys-2}
    \begin{array}{l}\displaystyle
      \sigma'= 3\sigma(\sigma-1)\,\frac{1-\tau^2}{1+\tau^2},
      \\[3mm]
      \displaystyle
      \tau'=-\frac{3}{2}(1-\tau)^2, \\
    \end{array}
\end{equation}
which can be solved explicitly. Indeed, the integration for $\tau$
results in
\begin{equation}\label{tau}
    \tau= 1+\frac{2}{\mathfrak{N}},
\end{equation}
where $\mathfrak{N}=3(N-N_0)$, and $N_0$ corresponds to some
initial data. Hence, $\tau\to 1$ at the end of evolution, i.e. at
$N\to +\infty$. Then,
\begin{equation}\label{tau1}
    \begin{array}{rcl}\displaystyle
    {\rm i}\ln\frac{1-\sigma}{1-\sigma_0}\,\frac{\sigma_0}{\sigma}&=&
    \displaystyle
    \ln\frac{\mathfrak{N}-\mathfrak{N}_c}{\mathfrak{N}-
    \mathfrak{N}_c^*}+
    \mathfrak{N}_c\ln(\mathfrak{N}-\mathfrak{N}_c)
    \\[4mm] && \displaystyle -
    \mathfrak{N}_c^*\ln(\mathfrak{N}-\mathfrak{N}_c^*),\\[1mm]
    \end{array}
\end{equation}
where $\sigma_0<1$, $\mathfrak{N}_c=-(1+{\rm i})$. Therefore, at
$N\to+\infty$ we find
$$
    \ln\frac{1-\sigma}{1-\sigma_0}\,\frac{\sigma_0}{\sigma}\to -2\ln N\to
    -\infty,
$$
that implies
\begin{equation}\label{tau2}
    \sigma\to 1,
\end{equation}
and the attractor is posed at the boundary circle,
\begin{equation}\label{tau3}
    x_\star=y_\star=\frac{1}{\sqrt{2}}
    \quad\mbox{or}\quad
    x_\star=y_\star=-\frac{1}{\sqrt{2}}.
\end{equation}

If the initial data $\sigma_0=1$, the quantity $\sigma$ does not
evolve, while $\tau$ approaches the attractor $\tau_\star=1$.

In addition, we have explicitly found that the second critical
point $\sigma_\star=0$, i.e. $x_\star=y_\star=0$ is unstable.

\begin{figure}[ht]
  \includegraphics[width=7.5cm]{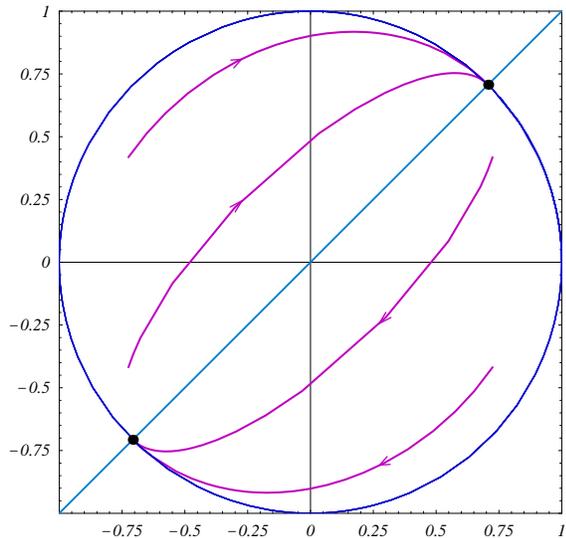}\\
  \caption{The attraction of trajectories to the critical points
  at the boundary circle of $\{x,y\}$ plane in the degenerate case.}\label{degenerate}
\end{figure}

The character of attractor (\ref{tau3}) is illustrated in Fig.
\ref{degenerate}, where trajectories move clockwise.

Thus, attractors (\ref{tau3}) exhibit the stable behavior in
appropriate semicircles, while the line connecting the critical
points has the instability versus perturbations.

\section{Phenomenological points}
\subsection{The mass}
The potential of quintessence suggests the mass
\begin{equation}\label{mass1}
    m^2=\frac{\partial^2 V}{\partial\phi^2}\Big|_{\phi=0}=
    4\pi G\lambda^2\widetilde V_0.
\end{equation}
For the exact scaling solution in Section II we get
\begin{equation}\label{mass2}
    m^2_s=\frac{9}{4}\,(1-w_B^2)\,H_0^2\Omega_\Lambda,
\end{equation}
and at $w_B=0$ and $\Omega_\Lambda\approx 0.7$ of practice the
mass is determined by the current value of Hubble constant, i.e.
it is extremely small as well as the energy scale of cosmological
constant, that is beyond a natural reason. Although, such the mass
could argue for both the present acceleration in the universe
expansion and the scale of cosmological constant.

Generically, we get
\begin{equation}\label{mass3}
    m^2=\frac{2\pi}{3}\,V_0 G\,z_\star^2=\frac{1}{4}\,
    H_0^2\Omega_\Lambda\,z_\star^2.
\end{equation}
So, the mass of scalar quintessence scales as the present day
Hubble constant with the factor of $z_\star$, which could be
arbitrary to enlarge the mass up to reasonable values in the
physics of Standard model. However, huge values of $z_\star$
involve extremely frequent oscillations of quintessence in the
nearest future, that is in contradiction with the present smooth
evolution of universe. Therefore, we expect that a viable model
includes the quintessence mass of the order of Hubble constant
today.

\subsection{Restriction to the slope}

The late time scaling of quintessence results in a fixed fraction
of quintessence energy in the budget of universe with respect to
other matter irrespective of the evolution stage: the dust or
radiation fix close values of fractions. However, the fraction of
nonbaryonic matter is constrained due to measured and primordial
abundances of light elements caused by Big Bang Nucleosynthesis
\cite{Wetterich,CLW}. Then, the slope of potential should be quite
large to suppress $\Omega_\phi\leqslant 0.13$, so according
\cite{CLW} one gets
\begin{equation}\label{lam10}
    \lambda^2>20.
\end{equation}

Next, a role of quintessence field $\phi$ during inflation
actually was analyzed in \cite{CLW}, since the hyper-sine in fact
coincides with the exponential potential at large values of field.
The problem is a relic abundance of quintessence after inflation,
that should be small in order to conserve the standard scenario of
nucleosynthesis. Appropriate restrictions in various schemes of
inflation are given in \cite{CLW}.

\subsection{Initial conditions}

\begin{figure*}
  \includegraphics[width=7.25cm]{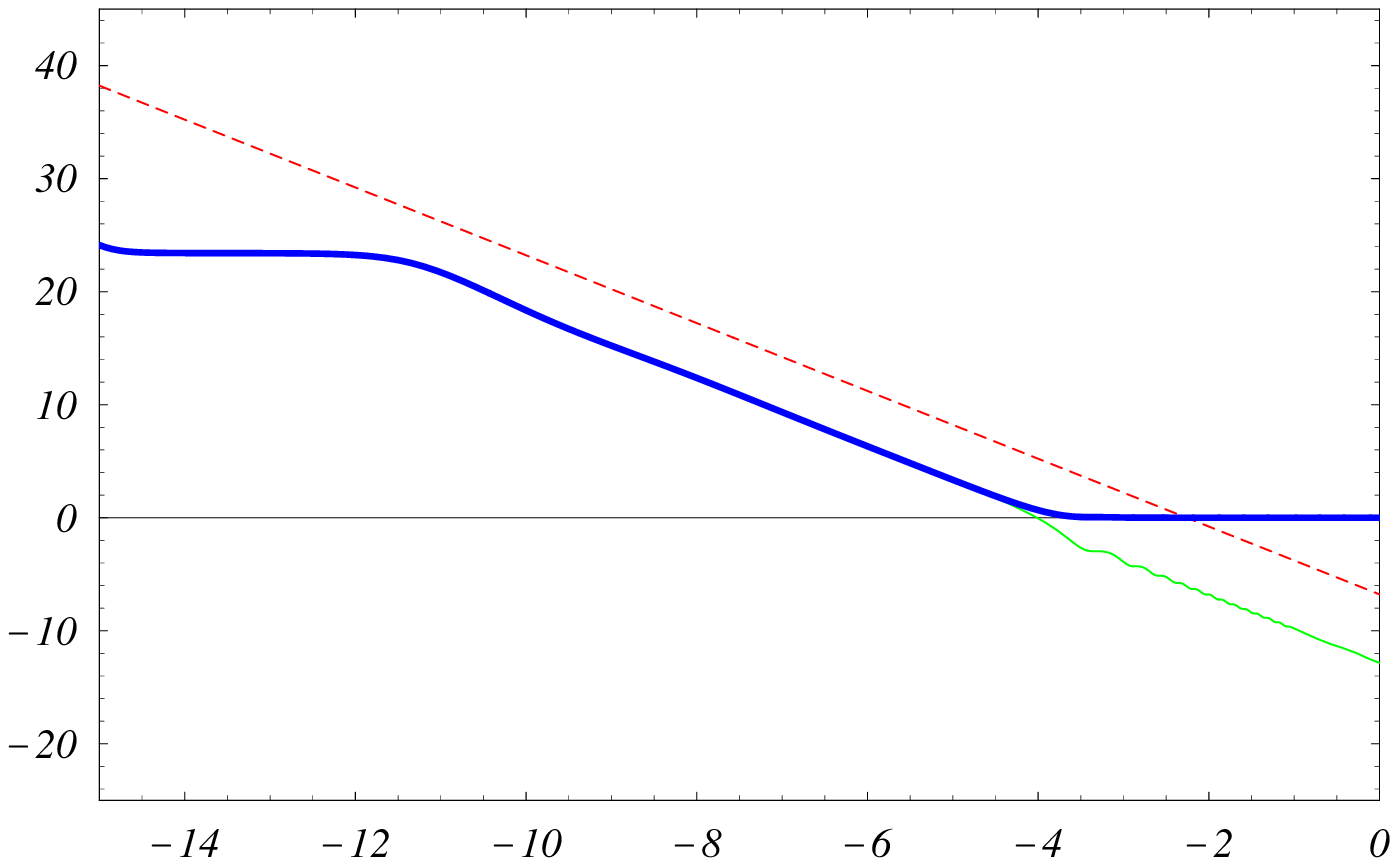}
  \hspace*{-1cm}\raisebox{4cm}{a)}
  \hspace*{7.5mm}
  \includegraphics[width=7.25cm]{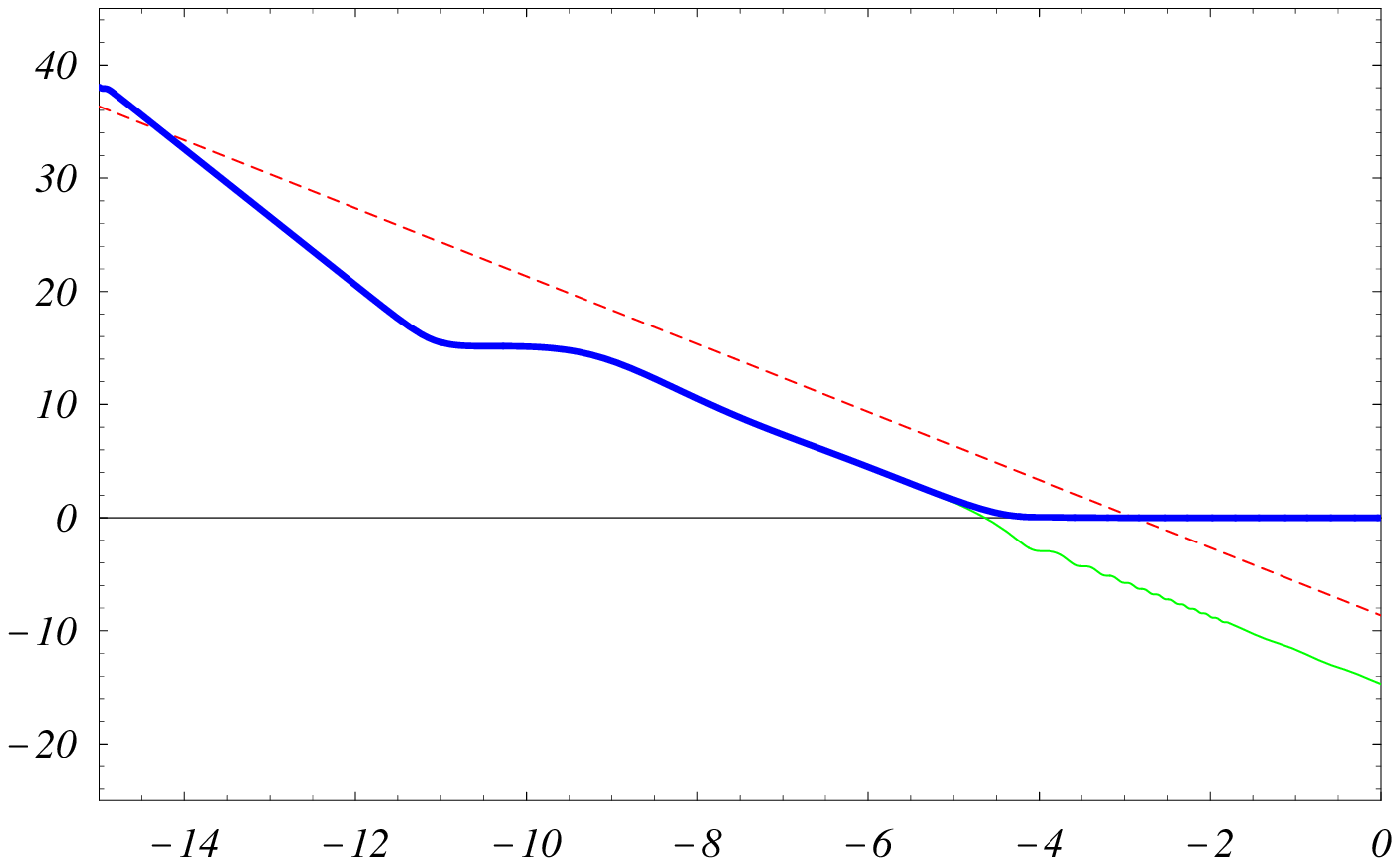}
  \hspace*{-1cm}\raisebox{4cm}{b)}\\[5mm]
  \includegraphics[width=7.25cm]{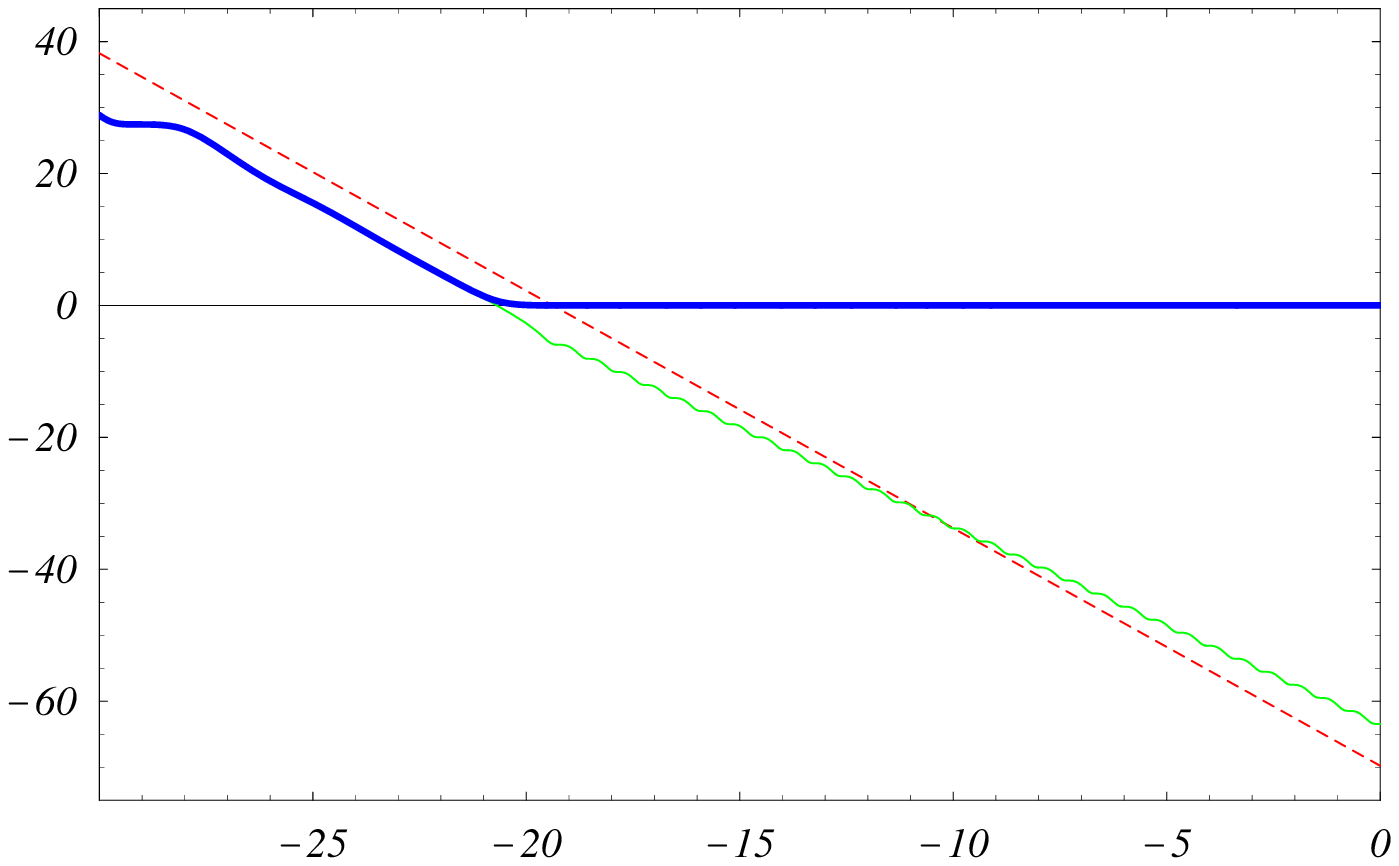}
  \hspace*{-1cm}\raisebox{4cm}{c)}
  \hspace*{7.5mm}
  \includegraphics[width=7.25cm]{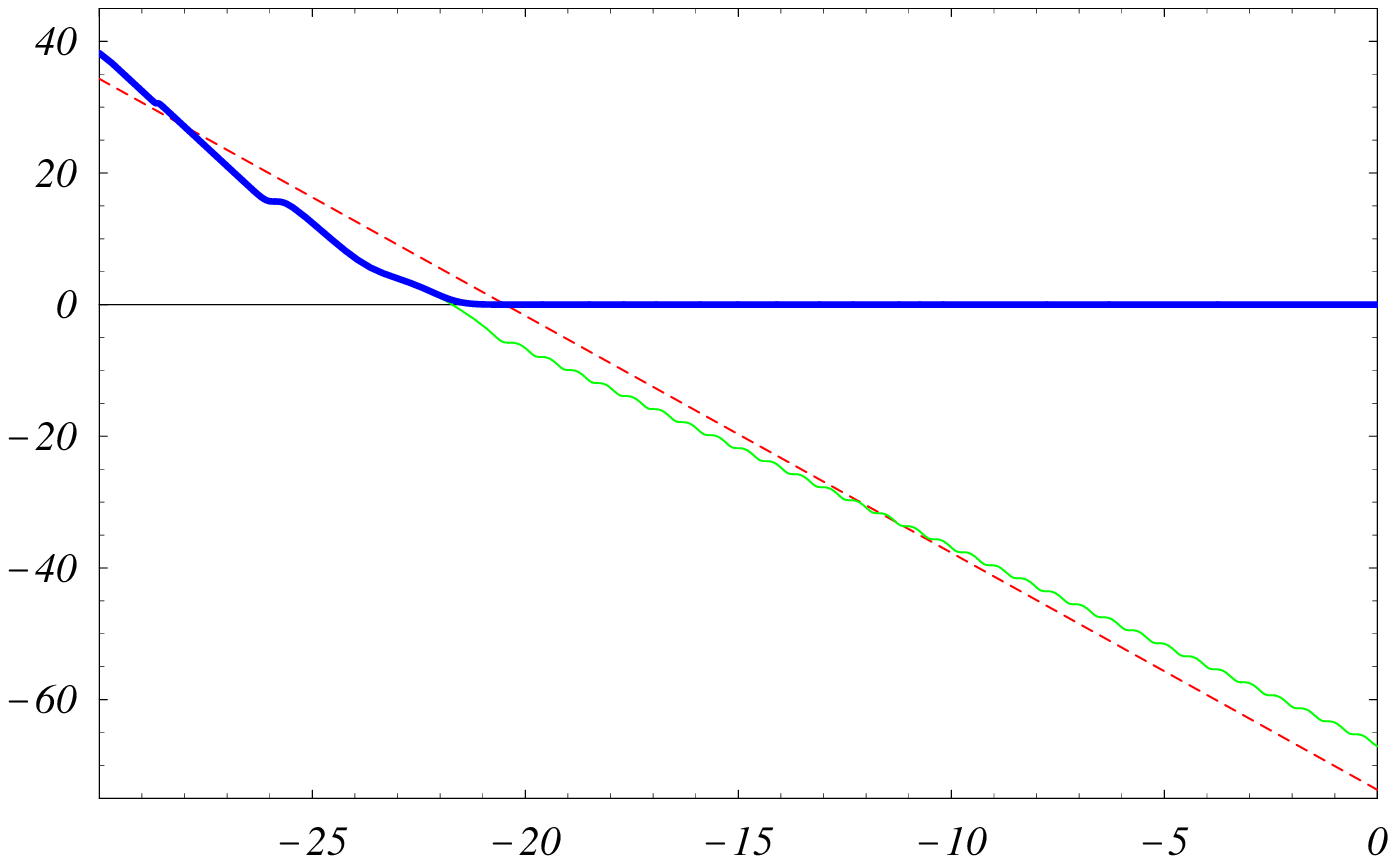}
  \hspace*{-1cm}\raisebox{4cm}{d)}\\[5mm]
  \includegraphics[width=7.25cm]{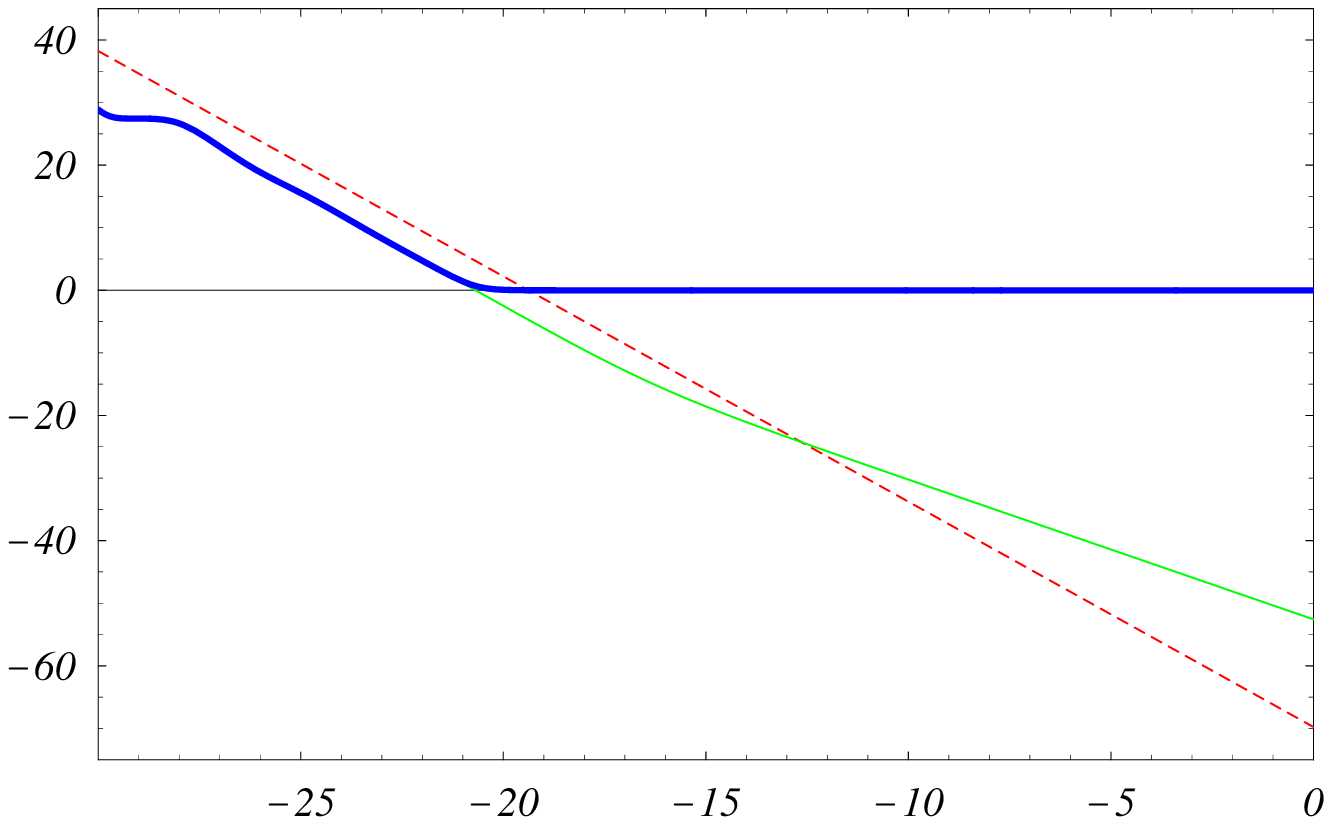}
  \hspace*{-1cm}\raisebox{4cm}{e)}
  \hspace*{7.5mm}
  \includegraphics[width=7.25cm]{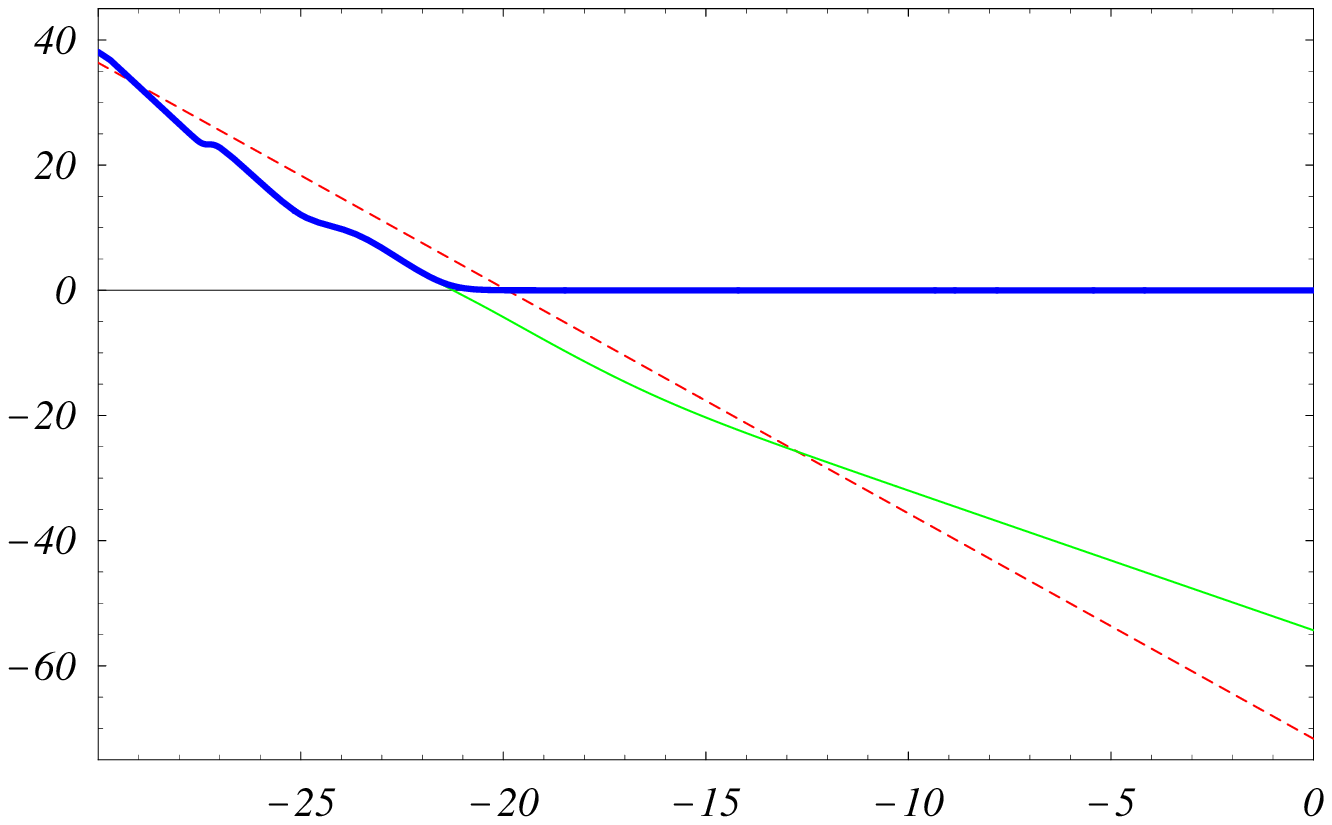}
  \hspace*{-1cm}\raisebox{4cm}{f)}\\[5mm]
  \includegraphics[width=7.25cm]{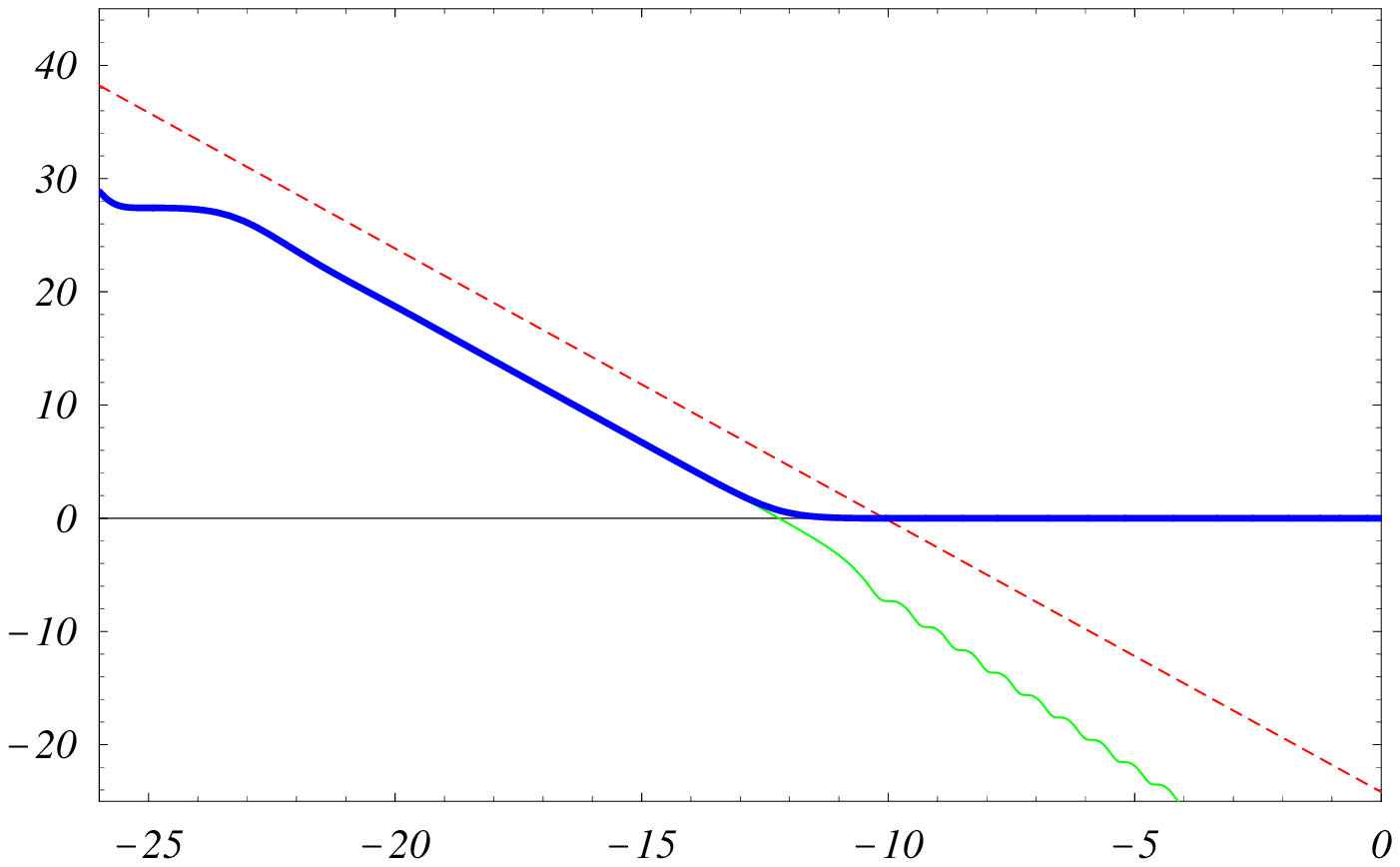}
  \hspace*{-1cm}\raisebox{4cm}{g)}
  \hspace*{7.5mm}
  \includegraphics[width=7.25cm]{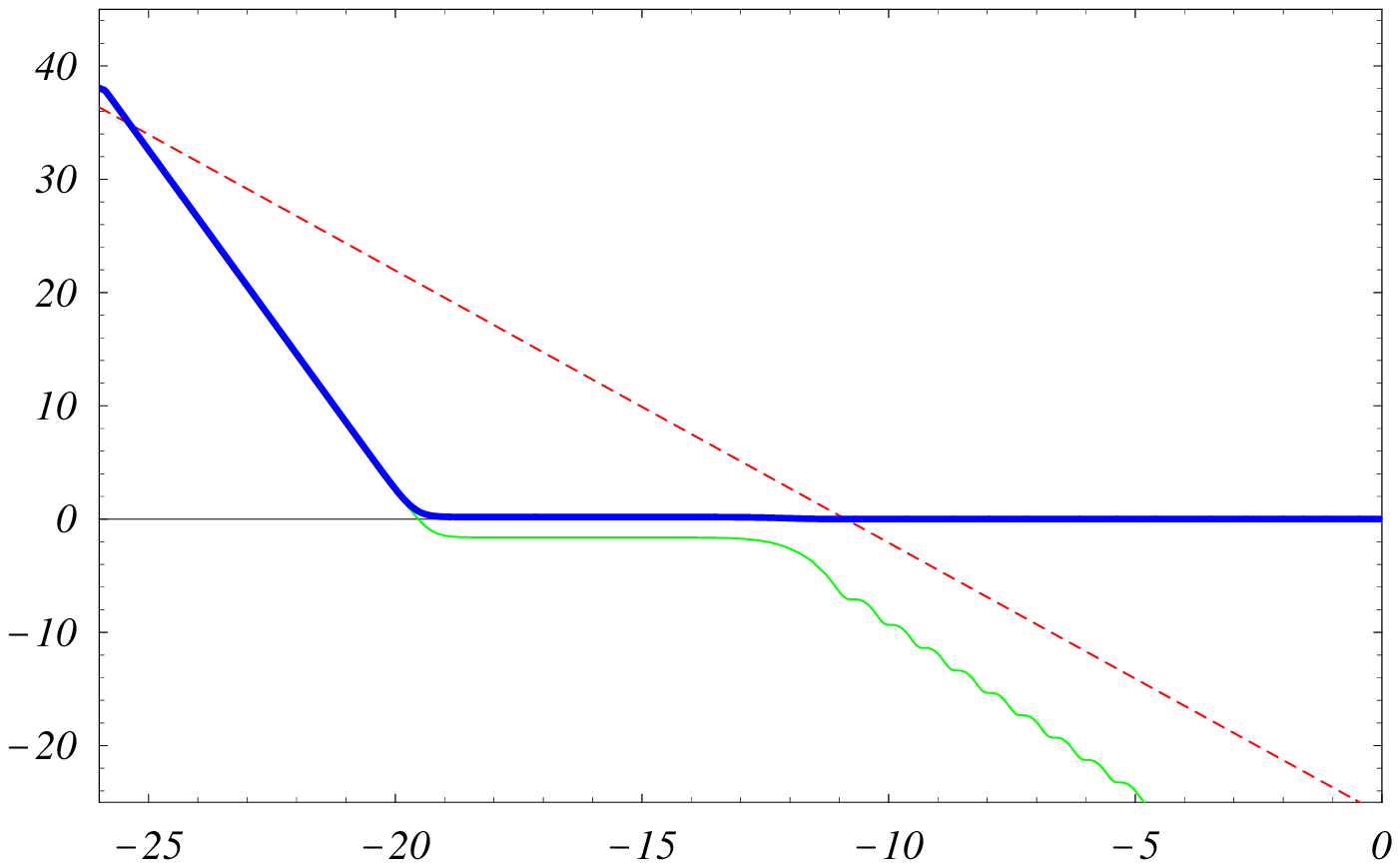}
  \hspace*{-1cm}\raisebox{4cm}{h)}\\[2mm]
  \caption{Approaching the scaling under different initial data:
  the logarithm of energy $\ln\rho$ in arbitrary units versus the
  e-folding \mbox{$N=\ln a(t)$} with a casual shift of present day scale factor $a(t_0)$;
  the quintessence and baryotropic matter
  are represented by thick solid and dashed lines, correspondingly, while
  the dynamical part of quintessence without the cosmological term is given
  by thin solid line. Initial data for the suppressed and dominated quintessence
  are respectively shown in left and right panels: a) and b) at $\lambda=20$, $w_B=0$
  and $z_\star=50$; c) and d) at $\lambda=20$, $w_B=0.2$ and $z_\star=10$;
  e) and f) at $\lambda=20$, $w_B=0.2$ and $z_\star=2.9$;
  g) and h) at $\lambda=20$, $w_B=-0.2$ and $z_\star=10$.}
  \label{falls}
\end{figure*}

Attractors mean a slow dependence of late time evolution on
initial data for the quintessence. The character of regulation is
illustrated in Fig. \ref{falls}. The set of tries exhibits the
following general features:
\begin{itemize}
    \item At small initial fraction of quintessence energy, it is
    frozen to a moment, when it approaches an appropriate scaling
    value in order to start the tracker behavior at late times.
    \item At large initial fraction of quintessence energy, it
    rapidly falls in order to frozen and wait for a moment of
    tracker way at late times.
    \item In future,  vibrations of quintessence at $z_\star>3$
    and $w_B\geqslant 0$ determine an average value of dynamical
    parameter for the equation of state $\langle\widetilde w_\phi\rangle$,
    which is independent of initial data, whereas $-1<\langle\widetilde w_\phi\rangle
    <w_B$ at $w_B>0$ or $\langle\widetilde w_\phi\rangle=0$ at
    $w_B=0$ (see Fig. \ref{falls} a, b, c, d). At $w_B<0$ vibrations determine the effective
    $\langle\widetilde w_\phi\rangle>w_B$ (see Fig. \ref{falls} g, h).
\end{itemize}

Further observations repeat general properties of future
attractors.

\subsection{Equation of state}

\begin{figure}
  \includegraphics[width=7cm]{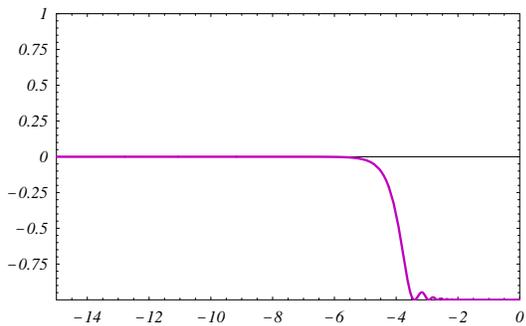}\\
  \caption{The state parameter of quintessence $w_\phi$ versus the
  e-folding $N$ of evolution scale: changing the stable value of late times
  at $w_B=0$ at $z_\star=50$ and $\lambda=20$.}\label{wphi}
\end{figure}

At late times, the attractor causes the ratio of quintessence
pressure to the energy density $w_\phi$ is stabilized infinitely
close to the value of parameter $w_B$ for the matter. However, the
situation changes, when the cosmological term comes to dominate,
and $w_\phi$ moves to $-1$. An example of $w_\phi$ relaxation is
shown in Fig. \ref{wphi} for the quintessence vibrating around the
minimum point (see Figs. \ref{phase-plot} f and \ref{falls} a, b).
The present day e-folding is arbitrary in Fig. \ref{wphi}. The
magnitude of deviation from the limit of $-1$ and vibration period
depend on the potential parameters. The picture analogous to Fig.
\ref{wphi} was observed in \cite{BarreiroCN} with similar values
of parameters.

It is clear that vibrations are absent if $z_\star<3$. Anyway, the
parameter of quintessence state $w_\phi$ rapidly approaches the
vacuum value. Nevertheless, it would be interesting to see a
relative significance of quintessence with respect to the matter.
So, general consideration of attractors and Fig. \ref{falls}
demonstrate that the quintessence fraction can dominate or be
suppressed depending on the potential parameters, the value of
$z_\star$. It is evident that in the case of reaching the boundary
circle in the $\{x,y\}$ plane the effective, average value for the
state parameter is $\langle \widetilde w_\phi\rangle = 0$, while
the same value is clearly observed also in the case of $w_B=0$ and
$|z_\star|>3$ (see Fig. \ref{falls} a and b).

Thus, we get the definite understanding of phenomenological
properties for the evolution of quintessence with the specified
kind of potential in the presence of cosmological term.

\section{Conclusion}
 In this paper we have found the potential of scalar field quintessence, which
 gives the exact solution for the scaling evolution of flat universe
 in the presence of cosmological constant. The scaling behavior is
 consistent with the current empirical observations.

 We have investigated the stability of scaling behavior versus the
 variations in the slope and normalization of potential as well
 as in initial data. The
 analysis has revealed  two kinds of attractors. The late time
 attractor just before the cosmological constant is coming to
 play, is independent of normalization, and it is determined by the
 slope, that is consistent with the well-known result for the
 exponential potentials \cite{CLW}, representing the limit of
 large field for the potential found in the paper. The future
 behavior of quintessence under the dominance of cosmological
 constant depends on both the ratio of potential normalization to
 the vacuum energy density and slope in the special combination
 denoted by parameter $z_\star$. Generically, the future attractor
 differs from that of late time. So, the late time attractor reveals the
 strange behavior. We have classified the future attractors by
 their character and stability in linear analysis.
 The degenerate case of nonlinear dependence has been solved explicitly.
 Some phenomenological items have been considered, too.

 We conclude that analysis of scaling attractors can be useful for
 classifying the quintessence behavior at late times and in
 future.

 This work is partially supported by the
Russian Foundation for Basic Research, grant 04-02-17530.

\end{document}